\documentclass[preprint]{iucr}              
\usepackage{amssymb}
\usepackage{bm}
\usepackage[fleqn]{amsmath}
\usepackage{float}
\usepackage{url}
\usepackage{graphicx}
\usepackage{xcolor}
\usepackage[outdir=./]{epstopdf}
\usepackage{epsfig}
\usepackage{booktabs}

\newcommand{\SVI}[0]{$\bf{S^{6}}$}

\newcommand{\scalarsub}[2]{$#1_#2$}
\newcommand{\vdotv}[2]{${{\bf #1 \cdot #2}}$}

\journalcode{A}              

\begin{document}                  

	
	
	{\LARGE \emph{\today}} \\
	\title{SELLA - A Program for Determining Bravais Lattice Types}
	
	
	\cauthor[a]{Lawrence C.}{Andrews}{lawrence.andrews@ronininstitute.org}{}
	\author[b]{Herbert J.}{Bernstein}
	\author[c]{Nicholas K.}{Sauter}
	
	
	\aff[a]{Ronin Institute, 9515 NE 137th St, Kirkland, WA, 98034-1820 \country{USA}}
	\aff[b]{Ronin Institute, c/o NSLS-II, Brookhaven National Laboratory, Upton, NY, 11973 \country{USA}, Rochester Institute of Technology, c/o NSLS-II, Brookhaven National Laboratory, Upton, NY, 11973 \country{USA}}
	\aff[c]{Lawrence Berkeley National Laboratory, 1 Cyclotron Rd., Berkeley, CA, 94720 \country{USA}}
	
	
	\shortauthor{Andrews, Bernstein, and Sauter}
	
	
	
	
	\keyword{Unit cell reduction}
	\keyword{Delaunay}
	\keyword{Delone}
	\keyword{Niggli}
	\keyword{Selling}
	\keyword{Bravais Lattice}
	
	
	
	\maketitle                        
	
	\begin{synopsis}
		A method for determining likely Bravais lattice types based on Selling (Delone) 
		reduction. It is a complete, closed solution.
	\end{synopsis}
	
	\begin{abstract}
		We introduce a new Bravais lattice determination algorithm. SELLA is a 
		straight-forward algorithm and a program for determining Bravais lattice type 
		based on Selling (Delone) reduction. It is a complete, closed solution, 
		and it provides a clear metric of fit to each type.
	\end{abstract}
	
	{\bf Note:}  Boris Delaunay in his later publications used the Russian version 
	of his surname: Delone. We will follow that choice.\\

	
	\section{Introduction}
	
	We introduce a new Bravais lattice determination algorithm. The Bravais 
	lattice types were created by \citeasnoun{Bravais1850}. \citeasnoun{Delaunay1932} 
	and \citeasnoun{Niggli1928} developed methods for the identification 
	of the Bravais lattice type of a crystal using the measured unit cell 
	dimensions; their methods were exact only if the cell parameters 
	exactly corresponded to the actual type \cite{Patterson1957}. 
	Delone discussed the issues of scalars that have nearly zero values, 
	but the broader issues of other measurement errors were not 
	discussed. \citeasnoun{andrews2014} review the literature on 
	the efforts to create methods to utilize data that contain unavoidable 
	measurement error.
	
	SELLA is a algorithm and program for determining Bravais lattice 
	type based on Selling (Delone) reduction. It is a complete, closed solution, and it provides a clear metric of fit to each type.
	
	\section{Terminology}
	
\begin{figure}
			\caption{Bravais triangle with labeling as used in the International Tables}
			\includegraphics[width=3cm]{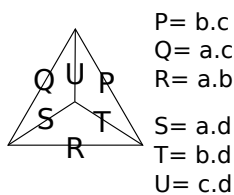}
			\label{fig:PQRSTU}
		\end{figure}

\begin{figure}
			\caption{Delone type description}
			\label{fig:terminology}
			\includegraphics[width=12cm]{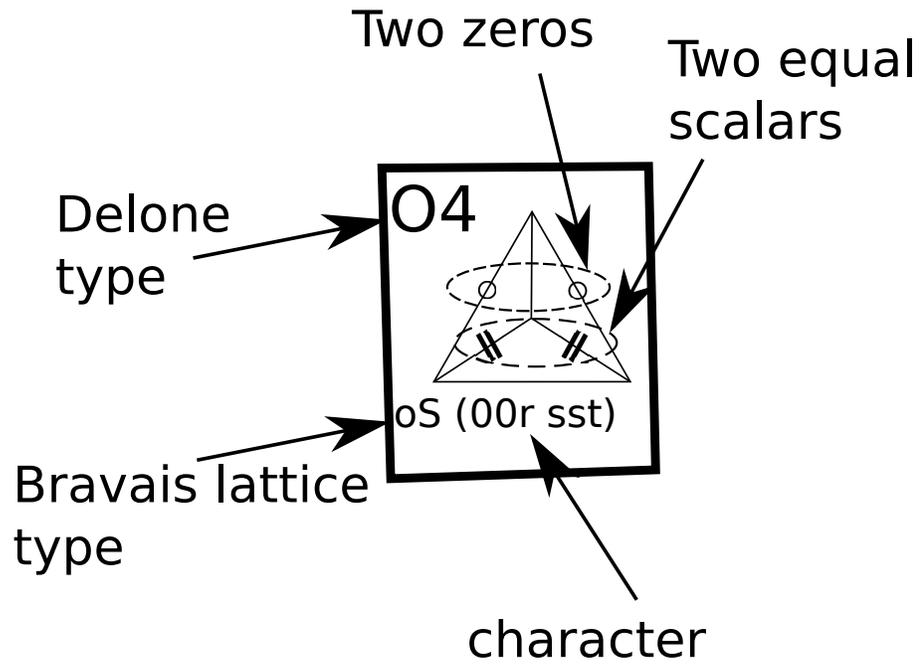}
		\end{figure}

	\citeasnoun{Delaunay1932} found that the 14 Bravais lattice types 
	could be allocated among 24 types that are determined by the 
	type of Dirichlet unit cell (termed the Voronoi region \cite{Voronoi1908} or 
	Dirichlet region \cite{dirichlet1850} by 
	Delone). Delone used the convention of \citeasnoun{Bravais1850}
	with the sides of the tetrahedron labeled by the labels of the six
	Selling scalars. Figure \ref{fig:PQRSTU} shows the labeling used in
	the International Tables for Crystallography \cite{Henry1952}; 
	Table \ref{table:ScalarConcordance} shows other choices that have been used. The
	contents for each type are described in Figure \ref{fig:terminology}.
	
	CHARACTER
	
\begin{figure}
			\caption{The Delone types}
			\label{fig:DeloneTypes}
			\includegraphics[width=12cm]{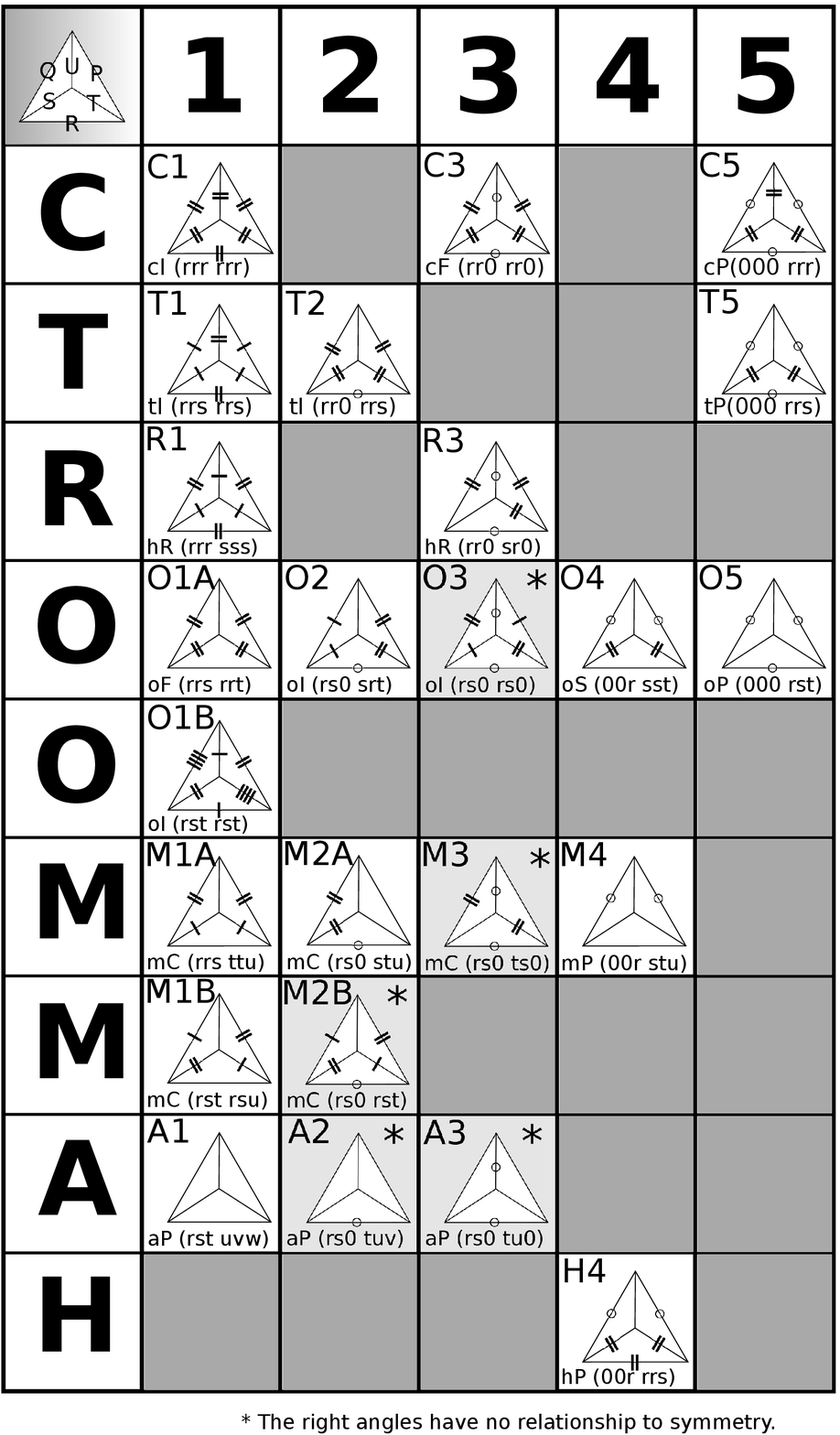}
		\end{figure}

	\citeasnoun{Burzlaff1985} revised the labeling of the types.
	Here we have returned to the numbering of the types to that 
	of \citeasnoun{Delaunay1932}.
	In some cases, Delone included two forms within a single 
	type. \citeasnoun{Burzlaff1985}
	renumbered by ordinals. Here we have labeled the Delone 
	types that have multiple items as``A'' and ``B''. Table \ref{table:DeloneTypeConcordance}
	shows the various labelings. Figure \ref{fig:DeloneTypes} 
	recapitulates the figure of \citeasnoun{Delaunay1932} using 
	more modern notation. Table \ref{table:ScalarConcordance} 
	lists the various labelings that have been 
	used for the sides of the Bravais tetrahedron. 
	Figure  \ref{fig:PQRSTU}  shows the arrangement of the 
	symbols as used in the International Tables for Crystallography.
	
	\begin{table}
		\begin{tabular}{cccccccccccc} 
			\toprule
			
			\rotatebox{80}{\cite{Delaunay1932}}
			& \rotatebox{80}{\cite{Burzlaff1992}}
			& \rotatebox{80}{\cite{Henry1952}  }
			& \rotatebox{80}{\cite{Patterson1957} }
			&\rotatebox{80}{\cite{andrews2019} } \\	
			\midrule
			g&p&P&h$_{23}$&\scalarsub{s}{1}\\		
			h&q&Q&h$_{13}$&\scalarsub{s}{2}\\		
			k&r&R&h$_{12}$&\scalarsub{s}{3}\\		
			l&s&S&h$_{14}$&\scalarsub{s}{4}\\		
			m&t&T&h$_{24}$&\scalarsub{s}{5}\\		
			n&u&U&h$_{34}$&\scalarsub{s}{6}\\		
			\bottomrule
		\end{tabular}
		
		\caption{The various terms used to describe the sides of 
			the Bravais tetrahedron. (In the figures in \citeasnoun{Delaunay1932} 
			the tetrahedron has been rotated by -120 degrees)}
		\label{table:ScalarConcordance}
	\end{table}

	\begin{table}
		\begin{tabular}{ccccccccc}
			\toprule
			\cite{Delaunay1932} & \cite{Burzlaff1985}   &this  \\
			&  & paper \\
			\midrule
			K$_I$ & K1 & C1 \\		
			K$_{III}$ & K2 & C3 \\		
			K$_V$ & K3 & C5 \\		
			Q$_I$ & Q1 & T1 \\		
			Q$_{II}$ & Q2 & T2 \\		
			Q$_V$ & Q3 & T5\\		
			R$_I$ & R1 & R1 \\		
			R$_{III}$ & R2 & R3 \\		
			O$_I$ & O1 & O1A  \\		
			O$_I$ & O2 & O1B  \\		
			O$_{II}$ & O3 & O2   \\		
			O$_{III}$ & O4 & O3   \\		
			O$_{IV}$ & O5 & O4   \\		
			O$_V$ & O6 & O5   \\		
			M$_I$ & M1 & M1A  \\		
			M$_I$ & M2 & M1B  \\		
			M$_{II_2}$  & M3 & M2A  \\		
			M$_{II}$ & M4 & M2B  \\		
			M$_{III}$ & M5 & M3  \\		
			M$_{IV}$ & M6 & M4  \\		
			T$_I$ & T1 & A1  \\ 		
			T$_{II}$ & T2 & A2 \\		
			T$_{III}$ & T3 & A3 \\		
			H$_{IV}$  & H1 & H4 \\ 
			\bottomrule
		\end{tabular}
		~\\ ~\\
		\caption{The 24 Delone types as labeled by various authors.}
			\label{table:DeloneTypeConcordance}
		\end{table}
		
		\subsection{The space \SVI{}}
		\citeasnoun{andrews2019} recast the Selling parameters to define a metric space,.
		We use the base vectors, $\bf{a}$, $\bf{b}$, $\bf{c}$ of the unit cell to define $\bf{d}=-(\bf{a}+\bf{b}+\bf{c})$.
		A point ${s}$ in \SVI{} is defined as:\\
		${s}$= [\scalarsub{s}{1}, \scalarsub{s}{2}, \scalarsub{s}{3},\scalarsub{s}{4}, \scalarsub{s}{5}, \scalarsub{s}{6}] \\
		where \scalarsub{s}{1}=\vdotv{b}{c}, ~\scalarsub{s}{2}=\vdotv{a}{c}, ~\scalarsub{s}{3}=\vdotv{a}{b}, 
		~\scalarsub{s}{4}=\vdotv{a}{d}, ~\scalarsub{s}{5}= \vdotv{b}{d}, and \scalarsub{s}{6}=\vdotv{c}{d}.
		
		\subsection{Terminology of the Bravais lattice types}
		The Bravais tetrahedron symbol is used to display the relationships among the \SVI{} scalars; see Figure \ref{fig:PQRSTU}  and Table \ref{table:DeloneTypeConcordance} for descriptions.
		The different ways that the scalars have been labeled is in  Figure \ref{fig:terminology}. 
		Figure \ref{fig:DeloneTypes} describes the information displayed for each Delone type.

		\section{Algorithms for creation of an exhaustive list of the polytopes of the Bravais types}
		
		\subsection{Deriving the projectors and perps}
		
		We will use Selling reduction; Niggli reduction has a fundamental 
		unit that is non-convex, and some boundaries of the fundamental 
		unit are partly closed and partly open \cite{andrews2014} leading 
		to considerable complexity. Because Niggli reduction divides the 
		fundamental unit into separate all obtuse and all acute regions, 
		the transitions between those are quite complex \cite{andrews2014}. 
		In space $\bf{S^{6}}$ \cite{andrews2019b}, the fundamental 
		unit is the simple all-negative orthant of a 6-dimensional 
		Cartesian space. Only two kinds of operations in the space 
		will need to be considered: the 6 boundary transform operations 
		at the boundaries of the orthant (zeros of an axis), and 
		24 reflections \cite{andrews2019}
		
		Twenty-four lattice types were enumerated by \citeasnoun{Delaunay1932}. 
		Three of those are triclinic, and we will ignore them because all crystals 
		can be indexed as triclinic. We 
		shall need all representations of all non-triclinic lattice 
		types within the $\bf{S^{6}}$ fundamental unit (all Selling-reduced 
		cells) or those generated beyond the boundaries by boundary transforms. 
		Three forms (M3, M2B, and O3) are reduced dimension boundaries that do
		not correspond to unique Bravais types; any lattice that falls in those
		three is indexed by one of the other 2 types or both. There remain 21, fully-6-dimensional,
		non-triclinic Delone lattice types.

		Seven of the 21 Delone types do not have a zero in their $\bf{S^{6}}$ 
		representation; the presence of zero indicates that a form is 
		on the boundary of the $\bf{S^{6}}$ fundamental unit. Since a 
		point on the boundary transforms to another boundary 
		point \cite{andrews2019b}, we can generate the Virtual 
		Cartesian Point (VCP) \cite{andrews2019b} appropriate to 
		each zero in the $\bf{S^{6}}$ vector. Because the reduction
		operations do not commute, in the cases where there is more 
		than one zero in the vector, all orderings of applying 
		reductions to all of the zeros must be generated. (For example, 
		for two zeros, there will be six vectors produced.) Again, 
		there may be many duplicate results. 
		
		Finally, the projectors must be computed for each of the many 
		sample vectors of the 21 types. We will also need the 
		``perps'', the operators that give the normal vector to each 
		such manifold. For a particular polytope, the projector applied to the probe gives
		the least-squares best fit point within that polytope. Perps are computed by subtracting the 
		corresponding projector from the unit matrix. The norm of a perp times
		a probe is the distance from that polytope.
		
		\subsubsection{Generating sample vectors from all Bravais types}

		Step 1: 		Generate any random vector that contains no zero 
		values and no duplicate values.  For a later operation, 
		it should only contain values large enough that adding or
		subtracting the smallest
		representable floating point number will not change the 
		value in computer arithmetic ($DBL\_MIN$ in the C language).
		
		Step 2: 
		For each Delone type, multiply its projector by 
		the vector from Step 1. Store  the results in a list 
		with entries for each type (see Table \ref{table:DeloneTypes} 
		for the projectors).
		
		Step 3: 	a) If the vector contains no zeros, generate the 24 reflections
		and store in the list by type. 
		
		Step 3: 	b) If the vector contains one zero, apply the boundary transform operation for that boundary
		to the vector, and
		store the products of the vector and its 24 reflections in the list. Note, 
		this is where the choice of $DBL\_MIN$ has effect; adding or
		subtracting such a small value will not change any value but zero. As a result, non-zero,
		integer values stored in floating point numbers will be unchanged.
		
		Step 3: 	c) If the vector contains more than one zero and \underline{not} $DBL\_MIN$, 
		first set the zero to $DBL\_MIN$,  and then return it to 
		step b). Then set the second zero to $DBL\_MIN$ (with the
		first zero still zero), and send that result to step b). 
		If there is a third zero,  process it the same way as two zeros and return to c).
		
		Continue to iterate those steps until all the types have been processed. 
		There will be duplicates within each type, which will be eliminated later.

		\subsubsection{Generate projectors from sample vectors}
		
		Each vector in the list from the previous steps is a representative of one
		of the Delone types. Although a vector is a one-dimensional polytope,
		these vectors encode the information about the full polytopes of the
		Delone types in their zeros and their duplicate values.
		
		We begin by defining a function: 
		\underline{Fraction} that computes the reciprocal of the 
		number of elements in the vector that equal 
		an input value or zero if there are no duplicates.
		
		The following steps will be repeated for each of the vectors in the
		list generated above. We generate a matrix \textit{m}.
		
		Step 1: Zero any scalars that have been set to $DBL\_MIN$. The result will
		be that all permutations will be created.
		
		Step 2: For each of the six scalars
		
		For i in 1-6
		\textit{m}[i,i] = \{if \textit{Fraction}(s[i]) == 0.0 or abs(s[i]) == 0.0\} then 0.0 else \textit{Fraction}(s[i])
		
		For i in 1-6
		For j in 1-6
		\textit{m}[i,j] 	= \{if s[i]==s[j] or abs(s[j]) $>$ 0\} then~\textit{Fraction}(s[j]) else 0.0
		
		Add m to a list of the projectors.

		To finish, the duplicate projectors are removed within each type, 
		leaving 239 non-triclinic projectors. (If the triclinic cases are 
		included, there are 10 more.) See Table \ref{table:TypeCounts}.

		\subsection {Performing the fit}
		The distance from the given polytope is just
		the norm of the perp for that polytope applied to the probe.
		
		\section{Degenerate Delone types}
		
\begin{figure}
			\caption{}
			\label{fig:degenerate}
			\includegraphics[width=12cm]{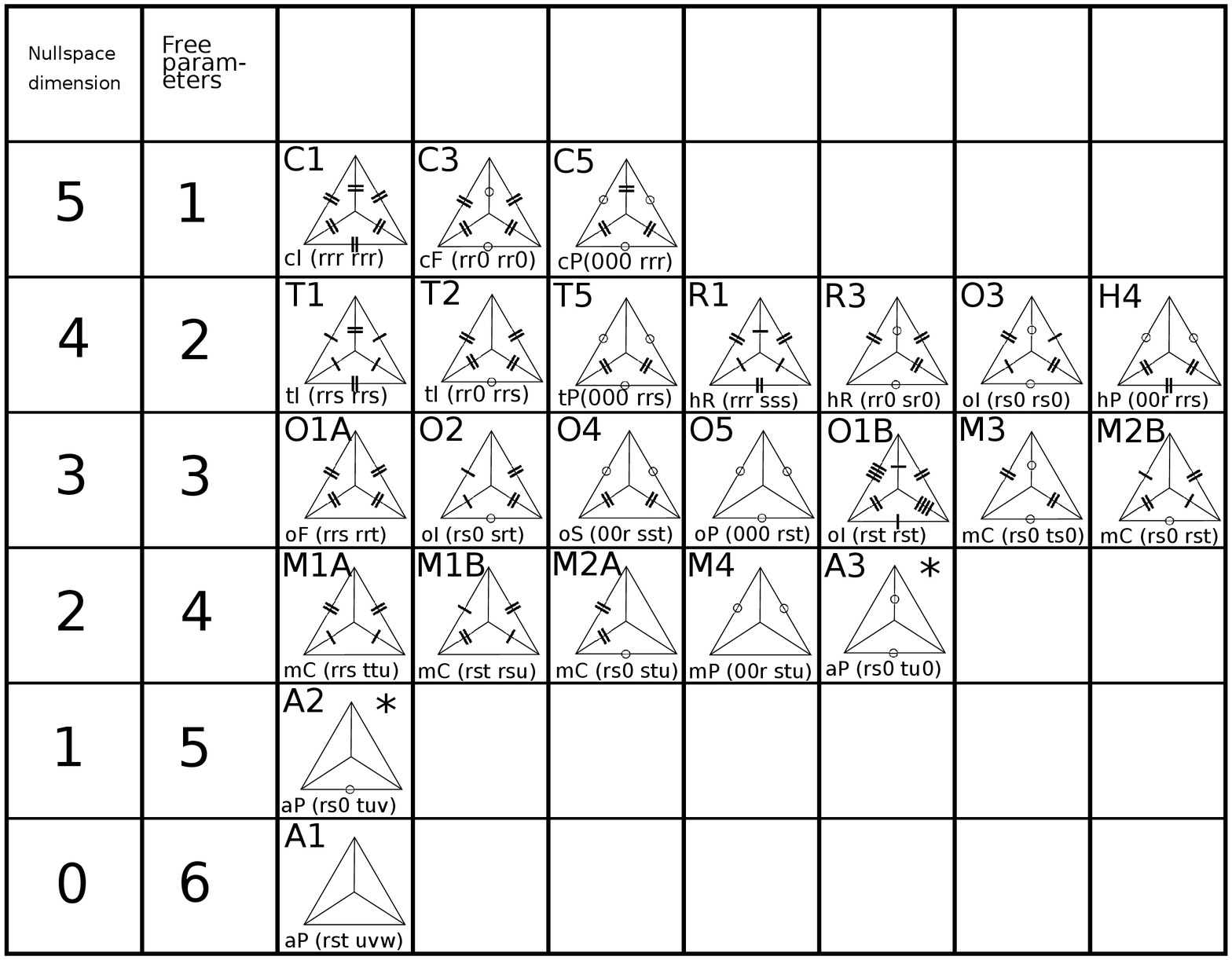}
		\end{figure}

		Three of the 21 non-triclinic Delone types are only boundaries of other types. In Figure 
		\ref{fig:degenerate}, Delone types M2B and M3 are in the row for orthorhombic, and
		O3 is in the row for rhombohedral and tetragonal. These types do not need to
		be searched for in general surveys.
		
		\section{Verification}
		
		\subsection{Rationale}
		
		Most of the proposals for Bravais lattice determination in the 
		presence of experimental error do not have a clear proof that 
		they are complete. For many years there were verbal complaints 
		that available programs would not infrequently miss experimental 
		cases (for example, TRACER \cite{lawton1965} and many anecdotal 
		communications).  To our knowledge, only the method of 
		\cite{OishiTomiyasu2012} has clearly demonstrated completeness. 
		The justification that Sella is complete has three parts.
		
		First: We require that the cell being tested be reduced, and
		all Bravais lattice types within the \SVI{} fundamental unit (all
		\SVI{} scalars negative) be generated. Therefore,
		the 24 possible forms of a 
		reduced cell will be close to one of the 24 copies of the types. 
		For the reflections, we do not need to consider the case of 
		points near a boundary that might be outside; the boundary 
		transforms must be treated for that.
		
		Second: In the case of the boundary transform operations, we take the case of a point that is infinitesimally distant from the polytope of a 
		Bravais lattice type that has one zero in its scalars. Take 
		the point in the polytope that is closest to the experimental 
		point. We transform both by the corresponding boundary transform 
		operation. We have generated a new pair of points in a 
		different location, one of which is in some representation of 
		the same Bravais lattice type. The generated points are still 
		infinitesimally distant from each other.
		
		Third: Assume we have a point that is outside the fundamental 
		unit and is near an undiscovered polytope for a Bravais lattice 
		type. If the point is near a boundary defined by a single 
		zero, then we apply a boundary transform to the point and to 
		the Bravais polytope, bringing them both into the fundamental 
		unit. If all of the variations of all of the Bravais 
		lattice types within and on the fundamental unit have 
		been generated, then this putative undiscovered polytope 
		has already been included in the list of possibilities. 
		Similar arguments apply to the cases of two or three zeros.
		
		If we apply all of the boundary transform operations to the 
		representations of the Bravais lattice types, then 
		there are no places that a point in the fundamental 
		unit will fail to find the closest Bravais lattice 
		(or lattices). As stated above, there are 239 
		non-triclinic possible representations.
		
		What might occur if some reduced point has an unreduced 
		copy near the all-negative orthant?  Would it be 
		possible for it to be close to one of the copies of one 
		of the 21 Delone types? The corresponding reduced point is obviously in 
		the all-negative orthant. Further, because all possible 
		copies of all of the Delone type manifolds have been 
		generated, the reduced point will be near a copy of the 
		manifold it was near when it was unreduced.
		
		\subsection{Testing against known cells}
		
		89539 unit cells were extracted (April, 2019) from the Protein Data Bank 
		(\citeasnoun{bernstein1977} and \citeasnoun{berman2000}). 
		When evaluated with Sella, all but 57 were found at zero 
		distance from the described crystal class. The 57 were 
		identified as incorrect unit cells for the assigned 
		crystal class; however, the distances from the assigned 
		classes were small because the deviations from the 
		appropriate parameters were modest (see Table \ref{tab:PDBCells}).
		\begin{table}
			\label{tab:PDBCells}
			
			\begin{tabular}{ccccccc}
				\toprule
				PDB ID & Sp.Gr. & Delone type&distance ({{\AA}}) \\ 
				\midrule
				1NUI&P3121&H4&0.0439\\ 
				1NUI&P3121&H4&0.0439\\ 
				2NW8&P3121&H4&0.0401\\ 
				2J1L&P3121&H4&0.0307\\ 
				1VYI&P3121&H4&0.2504\\ 
				3FZ8&P32&H4&0.0575\\ 
				2J5W&P3221&H4&0.0543\\ 
				2VAM&P3221&H4&0.1166\\ 
				1Y8J&P3221&H4&0.0389\\ 
				1F4V&P3221&H4&0.0276\\ 
				2ZP9&P6&H4&0.0528\\ 
				1DDR&P61&H4&0.0726\\ 
				1DDS&P61&H4&0.0726\\ 
				1SGU&P61&H4&0.0295\\ 
				3CDC&P61&H4&0.0559\\ 
				2WAG&P6122&H4&0.0377\\ 
				1LBM&P6122&H4&0.0312\\ 
				2O9Z&P62&H4&0.0402\\ 
				1ELZ&P6322&H4&0.4035\\ 
				1SA0&P65&H4&0.0681\\ 
				3KQI&P6522&H4&0.0331\\ 
				2F9Y&P6522&H4&0.0470\\ 
				3E73&P6522&H4&0.0523\\ 
				3CAP&H3&H4&0.0586\\ 
				2UX6&H32&H4&0.0386\\ 
				1ODT&H32&H4&0.0816\\ 
				2FDI&P43&T5&0.0758\\ 
				3HQ7&P43212&T5&0.2869\\ 
				1E2Q&P43212&T5&0.5482\\ 
				\bottomrule
			\end{tabular}
			\begin{tabular}{ccccccc}
				\toprule
				PDB ID & Sp.Gr. & Delone type&distance ({{\AA}}) \\ 
				\midrule
				1SHN&P43212&T5&0.0491\\ 
				1C50&P43212&T5&1.9040\\ 
				2A1L&P43212&T5&0.0528\\ 
				1S2L&P43212&T5&0.0334\\ 
				1H9S&P41&T5&1.0149\\ 
				1H9R&P41&T5&1.0141\\ 
				2F0Q&P41&T5&0.0260\\ 
				3EYM&P41&T5&0.0476\\ 
				1WV5&P41212&T5&0.0250\\ 
				2DF3&P41212&T5&0.0387\\ 
				1L6O&P41212&T5&0.0346\\ 
				3A58&P41212&T5&0.2976\\ 
				2IU9&P41212&T5&0.0373\\ 
				4BTP&P41212&T5&0.0666\\ 
				1W54&P4212&T5&0.0420\\ 
				2PBE&P42212&T5&0.0548\\ 
				3BR5&P42212&T5&0.2203\\ 
				1GMD&P42212&T5&0.1400\\ 
				1GMC&P42212&T5&0.1402\\ 
				2CCN&P42212&T5&0.0447\\ 
				1CE1&P212121&O5&0.6472\\ 
				3DQ2&P212121&O5&0.1058\\ 
				1Z2A&P212121&O5&0.0670\\ 
				2HA3&P212121&O5&0.3070\\ 
				3BNJ&I4122&T2&0.9518\\ 
				4MGP&I-42d&T1&0.2128\\ 
				4ASL&C2221&O4&0.3383\\ 
				3K7M&P432&C5&0.0517\\ 
				3GBN&I213&C1&1.2661\\ 
				\bottomrule
			\end{tabular}
			
			\caption{Among the unit cell parameters extracted from the PDB, these 57 has cell parameters that did 
			not conform to the assigned crystal types. In all cases, the differences were modest to small.}
		\end{table}
		\subsection{Testing for continuity}
		
		The polytopes of the Bravais lattice types are characterized by required zeros
		of scalars and/or multiplets of equal values. For instance, all primitive 
		monoclinic cells have two 90 degree angles and so two zero scalars. Consider R1 as
		an example for multiplets. The character for R1 is $ \bf{(rrr sss)} $; two triplets.
		
\begin{figure}
			\label{fig:M4}
			\caption{Delone type M4: there are two zeros. A circle of 100 points with
				radius 0.1 is centered at 0,0. The lattices are reduced and the points are
				plotted at the distance from M4.} The red points indicate the original
			circle. The blue points are for the reduced cells.
			\includegraphics[width=12cm]{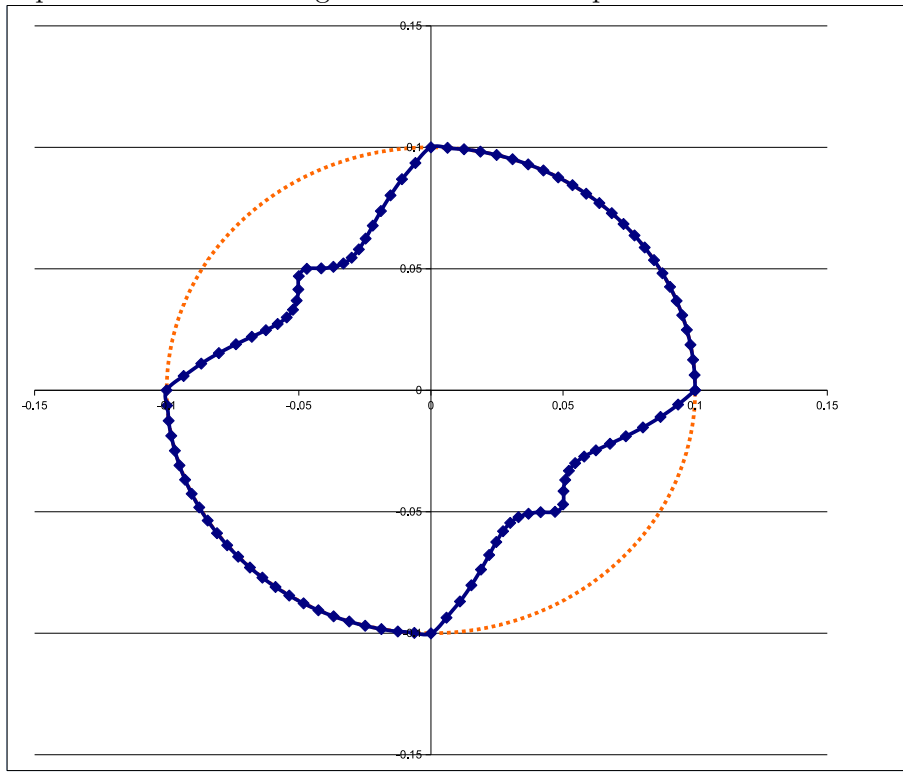}
		\end{figure}

		For the case of M4, primitive monoclinic, the character is $ \bf{(00r stu)} $. 
		For a particular value for r,s,t,u, the best monoclinic cell will not
		depend on whatever values are substituted for the zeros. In the plane
		corresponding to the two zero scalars, we can draw a circle, each point
		on the circle corresponding to a lattice. Reducing each so perturbed lattice and calculating
		the distance of the reduced cell from $ \bf{(00r stu)} $ will give a curve that
		should not have any discontinuities. See Figure \ref{fig:M4}. 
		
\begin{figure}
			\caption{Delone type O5: there are three zeros. A sphere 
				is centered at 0,0,0, with radius 0.1. The lattices are 
				reduced, and the point is 
				placed at a distance equal to the distance from type O5. 50,000
				points are plotted.}
			\label{fig:O5}
			\includegraphics[width=12cm]{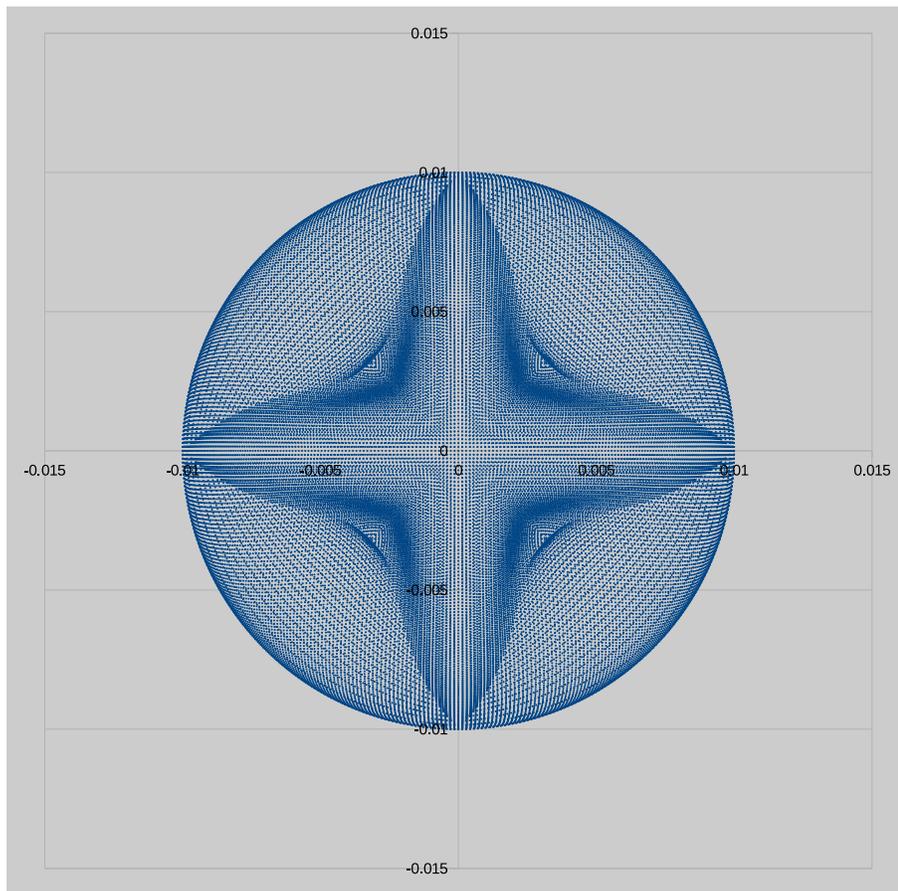}
		\end{figure}

		Similarly, Bravais type O5 has a triplet of zeros, character $ \bf{(000 rst)} $. Here we
		can generate a sphere that surrounds the orthorhombic point. See Figure \ref{fig:O5}.
		
\begin{figure}
			\caption{Delone type O4: There are two zeros and one pair.
				A sphere with radius 0.1 was placed around one point. At each
				sphere point, the lattice was reduced and the point on the sphere
				was assigned a color (0.01) to blue (0.0075). 100,000 points are plotted.}
			\label{fig:O4}
			\includegraphics[width=12cm]{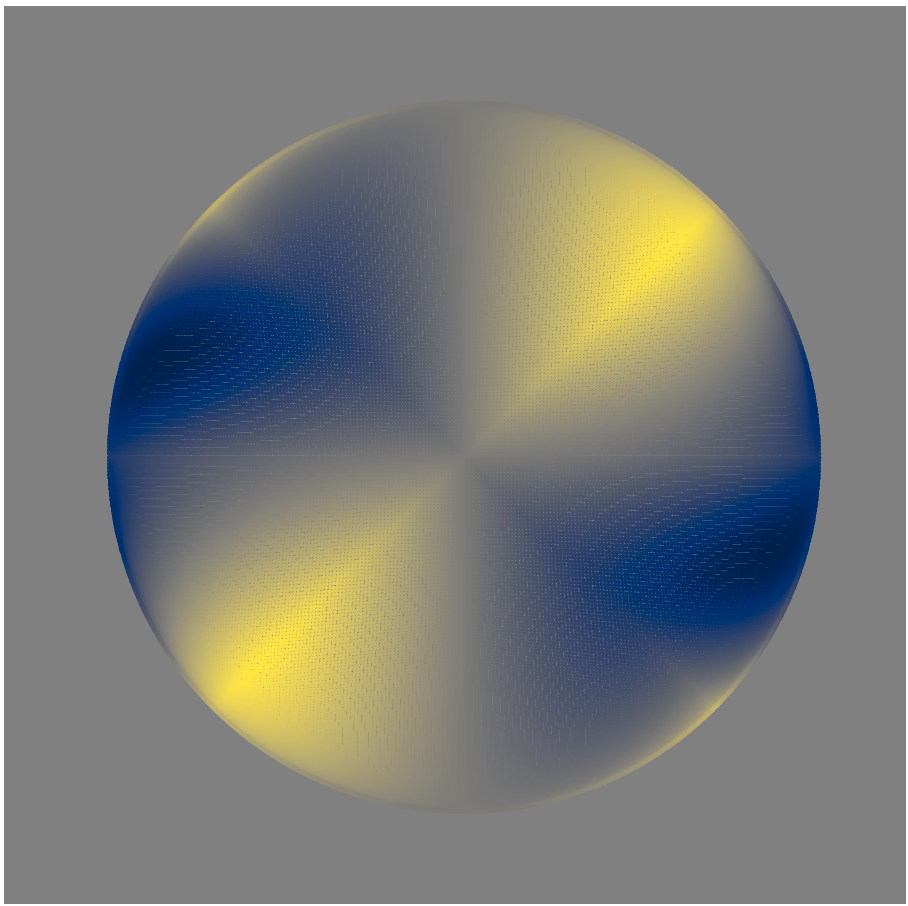}
		\end{figure}

		Bravais type O4 has two zeros and a pair of required equal values, 
		character $ \bf{(00r sst)} $. The zeros can be treated the same as above. 
		The pair of equal values must 
		have an average value that remains equal to $\bf{s}$ in the lattice character. 
		That leaves 3 values to vary, and we can create a sphere around 
		the point $ \bf{(00r sst)} $. See Figure \ref{fig:O4}.

		\section{Comparisons}
		
		Comparison to other distance measure; Data for other measures taken
		from \citeasnoun{andrews2014}.
		
		\begin{tabular}{ccccccc}
			\toprule
			Lattice character&G6 distance  &BGAOL   Z score&XDS QI &Scaled QI &Sella\\	
			\midrule
			mP               &     20.138  & 0.657    &   1.0  & 0.13  & 1.06\\	
			oC               &     125.958 &  3.560   &    23. & 8 3.09& 13.82\\	
			mC               &     125.150 &  4.085   &    23. & 4 3.03& 13.80\\	
			\bottomrule
		\end{tabular}
		
		~\\

		
		

		\ack{{\bf Acknowledgements}}
		Careful copy-editing and corrections by Frances C. Bernstein are 
		gratefully acknowledged.  	Our thanks to Jean Jakoncic and Alexei Soares for 
		helpful conversations and access to data and facilties at 
		Brookhaven National Laboratory.
		
		\ack{{\bf Funding information}}      
		
		Funding for this research was provided in part by: US Department of Energy Offices of Biological and Environmental Research and of Basic Energy Sciences (grant No. DE-AC02-98CH10886 ; grant No. E-SC0012704); U.S. National Institutes of Health (grant No. P41RR012408; grant No. P41GM103473; grant No. P41GM111244; grant No. R01GM117126, grant No. 1R21GM129570; grant No. P30GM133893); Dectris, Ltd.
		
		\bibliographystyle{iucr}
		\bibliography{Reduced}
		
		
		
		\begin{table}
			\caption{The 21 non-triclinic Delone types with the count of representations.
				The asterisks indicate non-crystallographic types.}
			\label{table:TypeCounts}
			\begin{tabular}{lr}
				Delone type & Count    \\
				\hline
				H4&12\\
				C1&1\\
				C3&3\\
				C5&16\\
				R1&4\\
				R3&12\\
				T1&3\\
				T2&6\\
				T5&48\\
				O1A&3\\
				O1B&1\\
				O2&6\\
				O3$^*$&9\\
				O4&36\\
				O5&16\\
				M1A&6\\
				M1B&3\\
				M2A&12\\
				M2B$^*$&12\\
				M3$^*$&18\\
				M4&12\\
				\bottomrule
			\end{tabular}
		\end{table}
		
		\fontseries{Courier}{
			\begin{table}
				\caption{Projectors for the 24 Delone types}
				\label{table:DeloneTypes}
				\begin{tabular}{lcl}
					\toprule
					C1  & (cI) & [1 1 1 1 1 1/  1 1 1 1 1 1/  1 1 1 1 1 1/  1 1 1 1 1 1/  1 1 1 1 1 1/  1 1 1 1 1 1] \\ 
					C3  & (cF) & [1 1 0 1 1 0/  1 1 0 1 1 0/  0 0 0 0 0 0/  1 1 0 1 1 0/  1 1 0 1 1 0/  0 0 0 0 0 0] \\ 
					C5  & (cP) & [0 0 0 0 0 0/  0 0 0 0 0 0/  0 0 0 0 0 0/  0 0 0 1 1 1/  0 0 0 1 1 1/  0 0 0 1 1 1] \\ \\
					
					T1  & (tI) & [1 1 0 1 1 0/  1 1 0 1 1 0/  0 0 1 0 0 1/  1 1 0 1 1 0/  1 1 0 1 1 0/  0 0 1 0 0 1] \\ 
					T2  & (tI) & [1 1 0 1 1 0/  1 1 0 1 1 0/  0 0 0 0 0 0/  1 1 0 1 1 0/  1 1 0 1 1 0/  0 0 0 0 0 1] \\ 
					T5  & (tP) & [0 0 0 0 0 0/  0 0 0 0 0 0/  0 0 0 0 0 0/  0 0 0 1 1 0/  0 0 0 1 1 0/  0 0 0 0 0 1] \\ \\
					
					R1  & (rP) & [1 1 1 0 0 0/  1 1 1 0 0 0/  1 1 1 0 0 0/  0 0 0 1 1 1/  0 0 0 1 1 1/  0 0 0 1 1 1] \\ 
					R3  & (rP) & [1 1 0 0 1 0/  1 1 0 0 1 0/  0 0 0 0 0 0/  0 0 0 1 0 0/  1 1 0 0 1 0/  0 0 0 0 0 0] \\ \\
					
					O1A & (oF) & [1 1 0 1 1 0/  1 1 0 1 1 0/  0 0 1 0 0 0/  1 1 0 1 1 0/  1 1 0 1 1 0/  0 0 0 0 0 1] \\ 
					O1B & (oI) & [1 0 0 1 0 0/  0 1 0 0 1 0/  0 0 1 0 0 1/  1 0 0 1 0 0/  0 1 0 0 1 0/  0 0 1 0 0 1] \\ 
					O2  & (oI) & [1 0 0 0 1 0/  0 1 0 1 0 0/  0 0 0 0 0 0/  0 1 0 1 0 0/  1 0 0 0 1 0/  0 0 0 0 0 1] \\ 
					O3  & (oI) & [1 0 0 1 0 0/  0 1 0 0 1 0/  0 0 0 0 0 0/  1 0 0 1 0 0/  0 1 0 0 1 0/  0 0 0 0 0 0] \\ 
					O4  & (oS) & [0 0 0 0 0 0/  0 0 0 0 0 0/  0 0 1 0 0 0/  0 0 0 1 1 0/  0 0 0 1 1 0/  0 0 0 0 0 1] \\ 
					O5  & (oP) & [0 0 0 0 0 0/  0 0 0 0 0 0/  0 0 0 0 0 0/  0 0 0 1 0 0/  0 0 0 0 1 0/  0 0 0 0 0 1] \\ \\
					
					M1A & (mC) & [1 1 0 0 0 0/  1 1 0 0 0 0/  0 0 1 0 0 0/  0 0 0 1 1 0/  0 0 0 1 1 0/  0 0 0 0 0 1] \\ 
					M1B & (mC) & [1 0 0 1 0 0/  0 1 0 0 1 0/  0 0 1 0 0 0/  1 0 0 1 0 0/  0 1 0 0 1 0/  0 0 0 0 0 1] \\ 
					M2A & (mC) & [1 0 0 1 0 0/  0 1 0 0 1 0/  0 0 0 0 0 0/  1 0 0 1 0 0/  0 1 0 0 1 0/  0 0 0 0 0 1] \\ 
					M2B & (mC) & [1 0 0 0 0 0/  0 1 0 1 0 0/  0 0 0 0 0 0/  0 1 0 1 0 0/  0 0 0 0 1 0/  0 0 0 0 0 1] \\ 
					M3  & (mC) & [1 0 0 0 0 0/  0 1 0 0 1 0/  0 0 0 0 0 0/  0 0 0 1 0 0/  0 1 0 0 1 0/  0 0 0 0 0 0] \\ 
					M4  & (mP) & [0 0 0 0 0 0/  0 0 0 0 0 0/  0 0 1 0 0 0/  0 0 0 1 0 0/  0 0 0 0 1 0/  0 0 0 0 0 1] \\ \\
					
					A1  & (aP) & [1 0 0 0 0 0/  0 1 0 0 0 0/  0 0 1 0 0 0/  0 0 0 1 0 0/  0 0 0 0 1 0/  0 0 0 0 0 1] \\ 
					A2  & (aP) & [1 0 0 0 0 0/  0 1 0 0 0 0/  0 0 0 0 0 0/  0 0 0 1 0 0/  0 0 0 0 1 0/  0 0 0 0 0 1] \\ 
					A3  & (aP) & [1 0 0 0 0 0/  0 1 0 0 0 0/  0 0 0 0 0 0/  0 0 0 1 0 0/  0 0 0 0 1 0/  0 0 0 0 0 0] \\ \\
					
					H4  & (hP) & [0 0 0 0 0 0/  0 0 0 0 0 0/  0 0 1 1 1 0/  0 0 1 1 1 0/  0 0 1 1 1 0/  0 0 0 0 0 1] \\ 
					\bottomrule
				\end{tabular}
			\end{table}
		}

	\end{document}